\documentclass[aps,amsmath,prl,12pt]{revtex4-1}
\usepackage{graphicx,color,bm,natbib,placeins}

\begin{document}
\hfill{\small Phys. Rev. Lett. {\bf 109}, 258701 (2012)~}

\title{Network Observability Transitions}

\author{Yang Yang$^1$, Jianhui Wang$^2$, and Adilson E. Motter$^{1,3}$}

\affiliation{$\mbox{1. Department of Physics and Astronomy, Northwestern University,}$ Evanston, IL 60208, USA  \\ $\mbox{2. Decision and Information Sciences Division, Argonne Natl Lab,}$ Argonne, IL 60439, USA\\
$\mbox{3. Northwestern Institute on Complex Systems, Northwestern University,}$ Evanston, IL 60208, USA}


\begin{abstract}

\bigskip
\noindent
In the modeling, monitoring, and control of complex networks, a fundamental problem concerns the comprehensive determination of the state of the system from limited measurements.   Using power grids as example networks,   we show that this problem leads to a new type of percolation transition, here termed a {\it network observability transition}, which we solve analytically for the configuration model.
We also  demonstrate a dual role of the network's community structure, which both facilitates optimal measurement placement 
and renders the networks substantially more sensitive to `observability attacks'.  Aside from their immediate implications for the development of smart grids, these results provide insights into decentralized biological, social, and technological networks. 
\end{abstract}


\maketitle 
Like other dynamical systems, a network is observable if its state can be determined from the given set of measurements, with observability depending on both the {\it number} and the {\it placement} of the measurements \cite{Monticelli1985}. This concept is important for a range of questions, including the identification of therapeutic interventions in intracellular networks, modeling and forecast in social networks, web crawling, 
monitoring and 
management of ecological networks, and  control of power-grid networks~\cite{network_review}.  Because measurements are inherently limited by cost and physical considerations, a question of interest concerns the identification of the optimal set of measurement points---e.g., sensors---with adequate redundancy that allow complete or (pre-specified) partial observability of the network. 

In  a power-grid network, the state of the system can be defined as the (complex) voltage at all nodes. Such state can in principle be determined by phasor measurement units (PMUs) \cite{PMU_book}, which are sensor devices that measure the voltage and line currents at the corresponding node in real time. Therefore, a PMU placed on a node makes 
both the node and (given the relation between current and voltage) all of its first neighbors observable---i.e., the states of those nodes are completely determined. 
If any of the neighboring nodes is a zero-injection node (i.e., without consumption or generation of power), then  a corresponding second neighbor may also be observable, and so on~\cite{PMU_Algthm_determin1}. In either case, the 
problem of identifying the observable nodes and the observability of the network itself is thus reduced to a purely topological one.

The observability of power-grid networks is a timely and broadly significant problem, which is also representative of many others.
Technologies that allow real-time wide-area monitoring of the network are an integral part of the next generation of power grids---the so-called smart grids \cite{Gungor_2011,Gellings_2004}---and PMUs are a central aspect of these technologies. 
It is believed, for example, that PMU information along with appropriate response could have prevented major recent blackouts \cite{2003_blackout_report}.  While the technology underlying PMUs is well established,  the high cost of required infrastructure, installation and operation continues to limit the number of such units that can be installed in a given power grid. Accordingly, significant recent research has been pursued in connection with PMU placement under various constraints for incomplete, complete, and redundant observability \cite{PMU_book}.  However, the fundamental question of how the observability of the network relates to its structure remains under-explored. 

In this Letter, we show that the random placement of PMUs leads to a new type of percolation transition \cite{Mendes2008}---a {\it network observability transition}. This transition characterizes
the emergence of macroscopic observable islands 
as the number of measurement nodes is increased. Using the generating function formalism \cite{newman_book}, we derive the exact analytical solution describing the size of the network's largest observable component (LOC).
We study its dependence on the network structure to show, in particular, how the transition threshold decreases as a function of both average degree and degree variance.  
We then consider the optimal placement of PMUs, a problem of practical interest that has been hindered  because no fast, deterministic algorithms currently exist to address large networks. Taking advantage of the community structure of real systems, we introduce a community-based  approach in which the network is judiciously partitioned into smaller, largely independent components that can be solved exactly. Our efficient approach allows us to address for the first time very large networks, including the largest interconnection of the North America power grid---a 56,892-node network.  We show, however, that community structure can also make the network 
more sensitive to the deliberate disabling  of PMUs, in that 
 a surprisingly small attack can separate the system into very small observable islands. 
 This adds a new dimension to existing research on the network vulnerability of power-grid systems \cite{Sole2008,Albert2004}.

We first consider random PMU placement on networks generated using the configuration model for a given degree distribution $P(k)$ \cite{cohen_book}. All nodes in the network are assumed to have a common probability $\phi$ of hosting a PMU. The observable nodes 
are classified into directly observable (hosting a PMU) and indirectly observable (neighboring a node with a PMU)
\cite{comment_1}.  
To determine the LOC size  as $\phi$ increases, we calculate the probability that a  randomly selected observable node $i$ is not connected to the LOC 
via a randomly selected edge  $e_{ij}$. 
This probability is denoted $u$ if node $i$ is directly observable and $s$ if neither $i$ nor $j$ is directly observable.
To proceed, we use
the generating function $G_0(x)=\sum_{k} P(k) x^k$, associated with the degree distribution, and the generating function $G_1(x)= G'_0(x)/G'_0(1)$, which describes the probability $q_\ell$ that, by following  a randomly selected edge, one reaches a node with $\ell$ other edges (i.e., $\ell$ excess edges) \cite{nsw_2001}.  

\begin{figure}[h] 
\includegraphics[width=11.5cm]{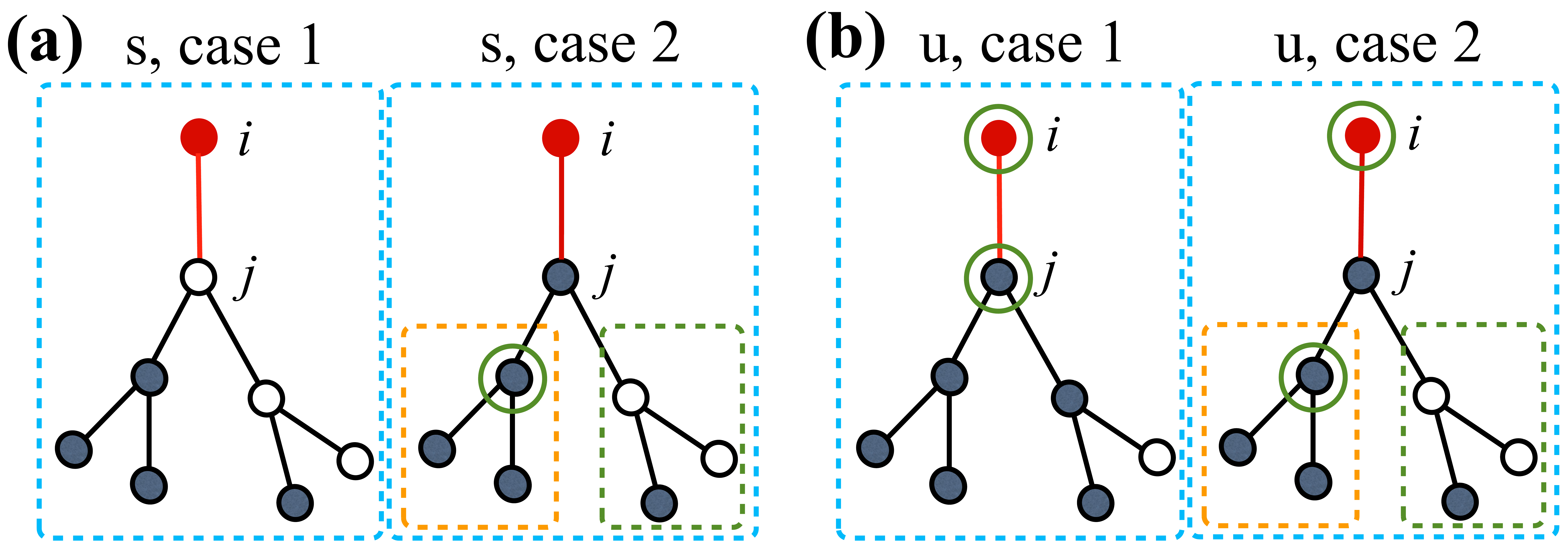}
\caption{Diagram for the self-consistent equations of (a) the probability $s$ and (b) the probability $u$ that an observable node $i$ (red circle) is not connected to the LOC through a specific edge $e_{ij}$ (red line).  The nodes are either not observable (open circles) or observable (solid circles), where green rings mark directly observable nodes.  
\label{fig2}}
\end{figure}

Self-consistent equations for the probabilities $s$  and $u$ can be derived as follows.  Starting with $s$, there are two independent cases in which node $i$ is not connected to the LOC via a randomly selected edge $e_{ij}$ [Fig.~\ref{fig2}(a)]. In the first case, node $j$ is not observable,
which occurs with probability  $G_{1}(1-\phi)$. In the second case, node $j$ is observable (i.e., $n\ge 1$ excess neighbors of  $j$ are directly observable) and hence the probability that an excess neighbor of node $j$ will not be connected to the LOC is $\phi G_{1}(u)$ if this neighbor is directly observable and $(1-\phi)s$ if it is not [Fig.~\ref{fig2}(a), orange and green subboxes]. Accounting for all possible degrees of node $j$ and values of $n$,  the latter case occurs with probability
$
\sum_{\ell=1}^{\infty}q_{\ell}
\sum_{n=1}^{\ell}
 {\ell \choose n}
\lbrack\phi G_{1}(u)\rbrack^{n}
\lbrack(1-\phi)s\rbrack ^{\ell-n}.
$
Combining both cases, we obtain the final expression for $s$:
\begin{equation}
\label{equ:s_selfconsis}
s = G_{1}(1-\phi) + G_1\left[  \Psi(s,u; \phi)    \right]- G_1\left[   (1-\phi)s   \right],
\end{equation}
where $\Psi(s,u; \phi) = \phi G_{1}(u) + (1-\phi)s$ corresponds to the probability that one indirectly observable node is not connected 
to the LOC via a specific edge. 

To derive a corresponding equation for $u$, we again split the problem into two cases [Fig.~\ref{fig2}(b)]. In the first case, node $j$ is directly observable but not part of the LOC, which occurs with probability $\phi G_{1}(u)$. In the second case, node $j$ is indirectly observable but not connected to the LOC via any of its excess edges, and this occurs with probability   $(1-\phi)G_1\left[\Psi(s,u; \phi)\right]$.
Combining these two cases, we arrive at the final expression for $u$:
\begin{equation}
\label{equ:u_selfconsis}
u = \phi G_{1}(u) + (1-\phi)G_{1}\left[\Psi(s,u; \phi)\right].
\end{equation}
Together, the self-consistent Eqs.~(\ref{equ:s_selfconsis})-(\ref{equ:u_selfconsis}) provide all the information needed to determine $s$ and $u$. 

\begin{figure}[t] 
\includegraphics[width=10cm]{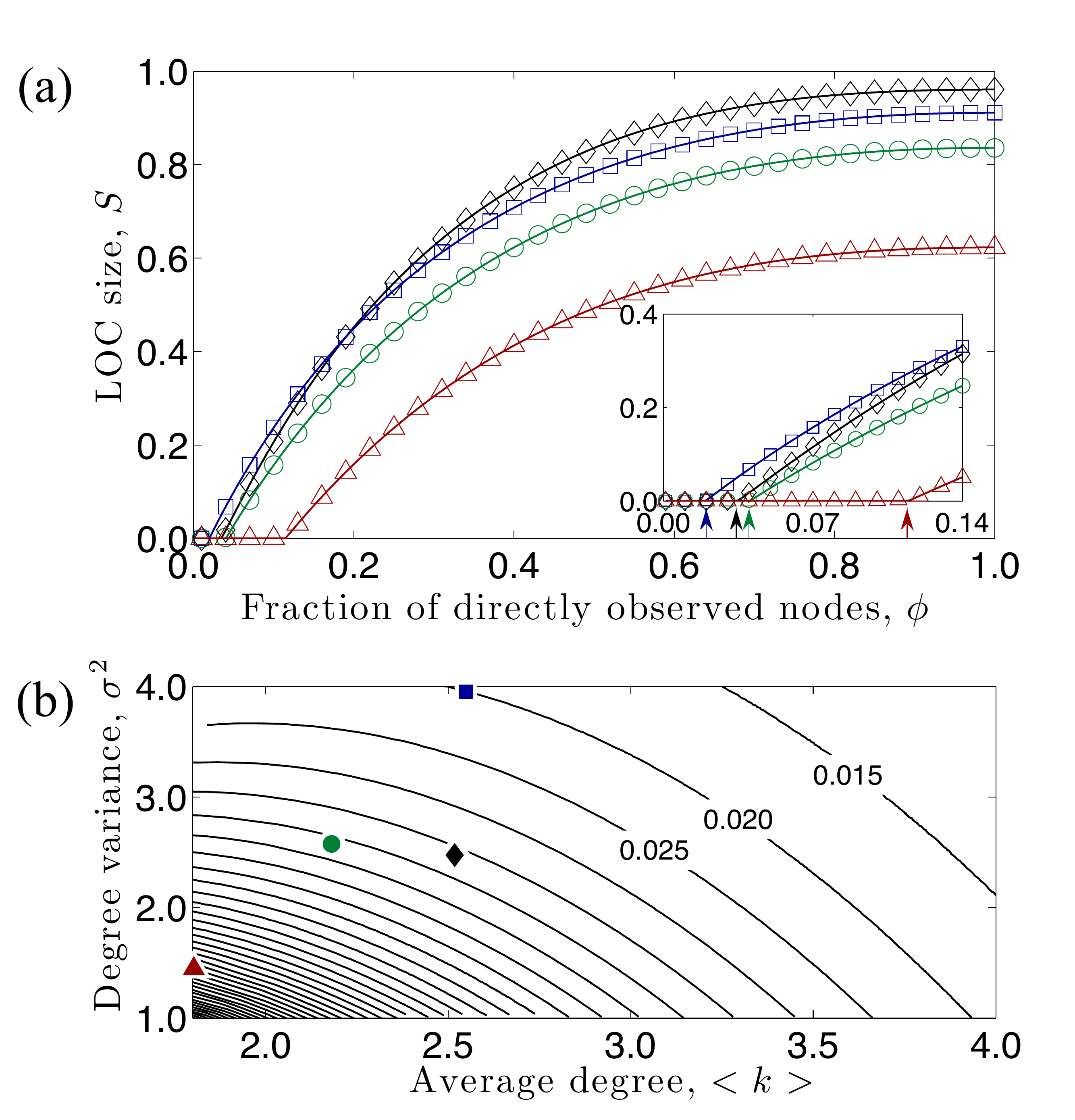}
\caption{Network observability transitions. (a)  LOC size as a function of $\phi$ in networks with the degree distributions of the power grids of Germany (red), Europe (green), Spain (blue), and Eastern North America (black) \cite{Sole2008,ferc}.  The continuous lines correspond to our analytical predictions, and the symbols to  an average over ten  $\mbox{$10^6$-node}$ random networks 
for $10$ independent random PMU placements each. The inset shows a magnification around the transitions, with the predicted thresholds $\phi_c$ 
indicated by arrows.  (b)  Dependence of $\phi_c$ on $\langle k\rangle$ and $\sigma^2$ for networks (in the thermodynamic limit) with Gamma degree distributions, where the curves indicate equispaced isolines of $\phi_c$ and the symbols indicate the $(\langle k\rangle, \sigma^2)-$positions of the corresponding networks in (a).
\label{fig3}}
\end{figure}

With $u$ and $s$  in hand, we now calculate the probability that a randomly selected node $i$ is part of the LOC.  If this node is directly observable,
which occurs with probability $\phi$, this probability is
$\sum_{k=1}^{\infty} P(k)\times (1-u^{k}) = \left[1 - G_{0}(u)\right]$.
This expression has the same form as for ordinary site percolation \cite{Callaway2000}, but 
here $u$ is functionally different. 
On the other hand, we also have to account for indirectly observable nodes.
If node $i$ is not directly observable, which occurs with probability $1-\phi$, this node is observable only if $m\ge 1$ of its neighbors are directly observable.  
Thus, the probability that node $i$ is part of the LOC is $1-G_{1}(u)^{m}s^{k-m}$, where the term $G_{1}(u)^m$ accounts for the $m$ directly observable neighbors and $s^{k-m}$ accounts for the $k-m$ other neighbors of $i$.  Considering all possible degrees $k$ of node $i$ and all possible values of $m$, the probability that this node is part of the LOC is 
$ \sum_{k=1}^{\infty} P(k) 
\sum_{m=1}^{k}  {k \choose m} 
\phi^{m}(1-\phi)^{k-m} \left[  1-G_{1}(u)^{m} s^{k-m} \right]$,
which can be rewritten as $1-G_{0}[\Psi(s,u; \phi)] - G_{0}(1-\phi) + G_{0}[(1-\phi)s] $.
Combining the two cases,  we obtain that the normalized size of the LOC is
\begin{eqnarray}
S = 1 - \phi G_{0}(u) - (1-\phi) \big\{G_{0}[\Psi(s,u; \phi)] 
				+ G_{0}(1-\phi) - G_{0}[(1-\phi)s] \big\}.
\label{equ:Giantcomsize}
\end{eqnarray}
This result is in excellent agreement with numerical simulations, as shown in Fig.~\ref{fig3}(a) for configuration-model networks with the degree distributions from a  selection of real power grids.

In particular, for a given degree distribution, and hence $G_{0}$ and $G_{1}$, there is a threshold $\phi_c$ at which $S$ becomes 
nonzero. This percolation threshold is given by 
 \small
\begin{equation}
\frac{(1-\phi)G'_1(1-\phi)}{G'_1(1)-1} \left[ 1-\phi G'_1(1)- \phi (1-\phi) G'_1(1)^2 \right] = 1,
\label{equ:phi_c} 
\end{equation}
\normalsize
which can be derived directly as the smallest $\phi$ at which Eqs.~(\ref{equ:s_selfconsis})-(\ref{equ:u_selfconsis}) hold for $s$ and $u$ smaller than $1$. It follows immediately from this expression that the threshold $\phi_c$ is strictly positive unless $G'_{1}(1)$ (and hence the second moment of the degree distribution) diverges. In power grids, however, the degree distribution is relatively homogeneous, meaning that an observability transition will occur at a nonzero value of the threshold $\phi_c$;  nevertheless, $\phi_c\ll 1$ for the degree distributions we consider.
This is emphasized in the inset of Fig.~\ref{fig3}(a), where the values of $\phi_c$ predicted in Eq.~(\ref{equ:phi_c}) are indicated by the arrows.

The threshold $\phi_c$ depends dominantly on the average degree $\langle k\rangle$ and the variance of the degree distribution $\sigma^2$.  The transition occurs earlier in denser and more heterogeneous networks, as illustrated in Fig.~\ref{fig3}(b). This diagram provides a 
close approximation to the positions of the transitions for the power grids shown in Fig.~\ref{fig3}(a), 
even though it was generated using Gamma degree distributions, which deviate from the approximately exponential distributions of the power-grid networks. 
This occurs because, even though $\phi_c$ can in principle depend on higher moments of the degree distribution through the term $G'_1(1-\phi)$,  this dependence is weak for systems with small $\phi_c$. 
We can show that for any degree distribution $\phi_c$ is upper bounded by a function $\phi_B$ of $G_1'(1) = \frac{\langle k^2\rangle -\langle k\rangle}{\langle k\rangle}$ that approaches zero rapidly as $G_1'(1)$ increases and, for fixed $G_1'(1)$, is lower bounded by a function $\phi_b$ of $\langle k^3\rangle/\langle k\rangle$ that decreases as this ratio increases (see supplement \cite{SI}).

These results provide insights relevant for real systems, but also point to other practical  considerations
concerning the observability of (necessarily finite-size and structured) real power-grid networks. For instance, what is the minimum number (and corresponding optimal placement) of PMUs needed for complete observability of an entire network? 

This optimization question can be formulated as a binary integer programming (BIP)  problem \cite{PMU_Algthm_determin1}, which is nevertheless NP-complete and hence not solvable in large networks. Meta-heuristic optimization methods can be relatively efficient \cite{PMU_complete3}, but the reliability of the solutions remains to be demonstrated.  Greedy algorithms \cite{PMU_Algthm_greedy}, on the other hand, are effective but provide only conservative estimates. A common feature of these approaches is that they do not take advantage of the internal organization of real power grids.  To proceed, we introduce an approach that is both efficient and effective. 
Specifically, we use modularity maximization \cite{Clauset2004} to split the network into communities, so that the placement problem within each community can be solved using BIP.  We solve the placement problem within one community, then we update the set of observable nodes on the whole network and move to the neighboring community most connected with the previously solved communities, and so on.  This procedure is repeated by starting from each of the communities; we select the minimum-PMU solution,  although for the systems considered here we verified that the community sequence has very small impact on the number of PMUs (e.g., relative standard deviation $< 2\times10^{-4}$ for the Eastern North America power grid).

\begin{table}[b] 
\caption{Optimal PMU placement based on community splitting, 
where $N$ is the number of nodes, $\langle k\rangle$ is the average degree, $Q$ is the modularity, 
$N_C$ is the number of communities, $N_P$ is the minimum number of PMUs estimated
from the community structure,  $N_P^{opt}$ is the exact minimum number (only computable for the small IEEE test systems), and $N_P^{g}$ is the  greedy optimal solution.
} 
\centering
\begin{tabular}{lrccrrcr}
\hline 
Power grid  
& $N$~ & $\langle k\rangle$~ & $Q$ & ~~$N_C$ & $N_P$~ & $N_P^{opt}$ &  $N_P^{g}$\\  
\hline \hline
IEEE118 & $118$~ & $3.16$~~ & $0.72$ & $8$ & $32$~ & $32$ & $36$\\
\hline
IEEE300 & $300$~ & $2.73$~~ & $0.83$ & $14$ & $89$~ & $87$ & $96$\\
\hline
PJM & $14,077$~ & $2.60$~~ & $0.95$ & $52$ & $4,246$~ & --- & $4,493$\\
\hline
Eastern & $56,892$~  & $2.52$~~ & $0.97$ & $96$ & ~$17,216$~ &--- & $18,216$\\
\hline
\end{tabular}
\label{table:results_comparison}  
\end{table}

As shown in Table \ref{table:results_comparison},  benchmarking of this approach using small networks that can be solved exactly shows that it offers
very good approximations of the optimal solutions.  
As a comparison, the application of  the greedy algorithm maximizing at each step the increase in
the fraction of observable nodes (FON) results in a solution that requires $1,000$ additional PMUs in the  Eastern North America power grid.  
For both this system and the PJM (Pennsylvania-New Jersey-Maryland)  power grid,  our approach shows that the resulting minimum number of PMUs for complete observability corresponds to approximately $30\%$ of the nodes in the network (Table~\ref{table:results_comparison}), which is comparable to previous estimates and exact calculations on small networks available in the literature \cite{WAMPS}.

\begin{figure*}[t] 
\includegraphics[width=6.5in]{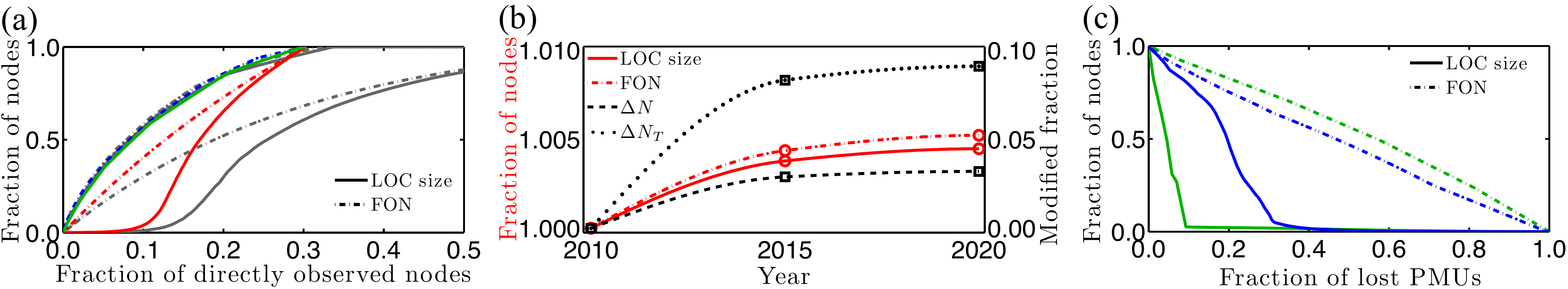}
\vspace{-0.3cm}
\caption{Observability on the largest available power grid 
(Eastern North America).
(a) Complete and incomplete observability: LOC size and FON for random PMU placement (red),
greedy LOC size optimization (green), and greedy FON optimization (blue) on the optimal set.
The corresponding curves for placement on the full network are shown in gray.
(b) Network evolution:  LOC size and FON on the planned networks for the years  2015 and 2020 given the 
 optimal PMU placement on the 2010 network (red). 
The black lines represent the net increase in the number of nodes ($\Delta N$) and the total number of nodes modified by
node additions, node removals, or edge-rewires ($\Delta N_T$).
To facilitate comparison, all curves are plotted relative to the initial number of nodes.
(c) Observability attack: LOC size and FON  for both FON attack (blue) and LOC size attack (green), 
where the latter
takes 
advantage of the community structure of the network.  
\label{fig4}}
\end{figure*}

Interestingly, an abrupt (albeit smooth) transition of the LOC size occurs also for real networks and even if we limit the random PMU placement to the {\it optimal set}   (i.e., the solution set of the optimal placement problem), as illustrated in  Fig.~\ref{fig4}(a)  (continuous red line).  The FON, in contrast, grows approximately linearly as the fraction of directly observed nodes increases from zero (dot-dashed red line). 
However, we can cause both the LOC size and the FON to grow sharply from the beginning by changing the placement sequence [Fig.~\ref{fig4}(a), green and blue lines, respectively]. 
Using optimal PMU placement on the 2010 Eastern North America power grid and data on the planned upgrades of the network until 2020 \cite{ferc}, we also demonstrate that both the LOC size and the FON are rather robust against the evolution of the network [Fig.~\ref{fig4}(b)]. Even after nearly 10\% of the nodes have been 
 removed, added, or rewired, 
  neither the LOC size nor the FON decreases (and they in fact increase) relative to the number of nodes in the initial network. 
(See supplement \cite{SI} for an analogous 
conclusion  when considering the impact of random edge-rewiring.) 
This suggests that an initially optimal (hence minimally redundant)  placement of PMUs  
remains effective as the network evolves. 

However, this does not mean that the network is robust against intentional `observability attacks', which we define as the deliberate disabling of PMUs (rather than of power-grid nodes themselves). In fact, while the FON remains large upon a sequential inactivation of PMUs that maximizes reduction of the FON at each step, the LOC size decreases rapidly (Fig.~\ref{fig4}(c), blue lines). Moreover, this decrease is significantly faster if we attack the LOC by targeting inter-community PMUs, effectively breaking the LOC into observable islands defined by the network community structure (Fig.~\ref{fig4}(c), green lines). Ironically, the same network property that facilitates identification of optimal PMU placement---community structure---makes the network vulnerable to observability attacks.

We suggest that similar analysis can also be useful for other networked systems, such as traffic monitoring in diverse networks and network discovery. For example, in content-based network crawling,  the initial nodes in the crawling problem play the role of directly observable nodes, and the emergence of a LOC indicates that a fraction of the nodes will be visited from multiple initial nodes. These problems invoke the notion of depth-$L$ observability, in which the direct observation of a node can make all neighbors within distance $L$ indirectly observable. While here we have focused on depth-1 observability, which is the most relevant for power-grid networks, our analysis can be extended to higher observability depths (see
 supplement \cite{SI} for a depth-2 example).  
These concepts can also  be extended to systems in which observability depends on additional network structural properties, as in the case of metabolic networks  (see supplement \cite{SI}).
Therefore, like other percolation processes studied previously \cite{Mendes2008,newman_book,cohen_book} and recently  \cite{eper1,eper2,eper3} on networks,  network observability transitions have implications for a wide range of systems.

Network observability is challenging in part because networks represent {\it distributed} dynamical systems, whose state cannot be assessed from single measurement points. However, randomness, long-range connections, and the consequent small node-to-node distances common to many real networks facilitate observability as they significantly reduce the necessary number of directly observable nodes. This underlies the finite but surprisingly small threshold for the observability transitions identified here even for fairly sparse and homogeneous networks. 
In infrastructure networks,  
wide-area observability and monitoring is  necessary for modernization of the systems'  
operation \cite{WAMPS}.
Yet, reliance on observability comes at the
risk of making  the network vulnerable to a new form of attack, in which the deliberate disabling of a relatively small number of sensors may render the network unobservable, 
hence potentially nonoperational, 
even when it  
is robust against conventional attacks \cite{Annibale2010}.

The authors thank Cong Liu for providing data and Yu Cheng for inputs on the data processing. 
This work was supported by a Northwestern-Argonne Early Career Investigator Award for Energy Research.

\newpage

\centerline{\Large Supplemental Material}
\vspace{0.5cm}

\noindent
Network Observability Transitions\\
Yang Yang, Jianhui Wang, and Adilson E. Motter \\

\renewcommand{\theequation}{S\arabic{equation}}

\setcounter{figure}{0}
\renewcommand{\thefigure}{{S\arabic{figure}}}

\noindent
\centerline{\bf Observability Transition Threshold in Heterogeneous Networks}

\begin{figure}
\centering
\includegraphics[scale = 0.75]{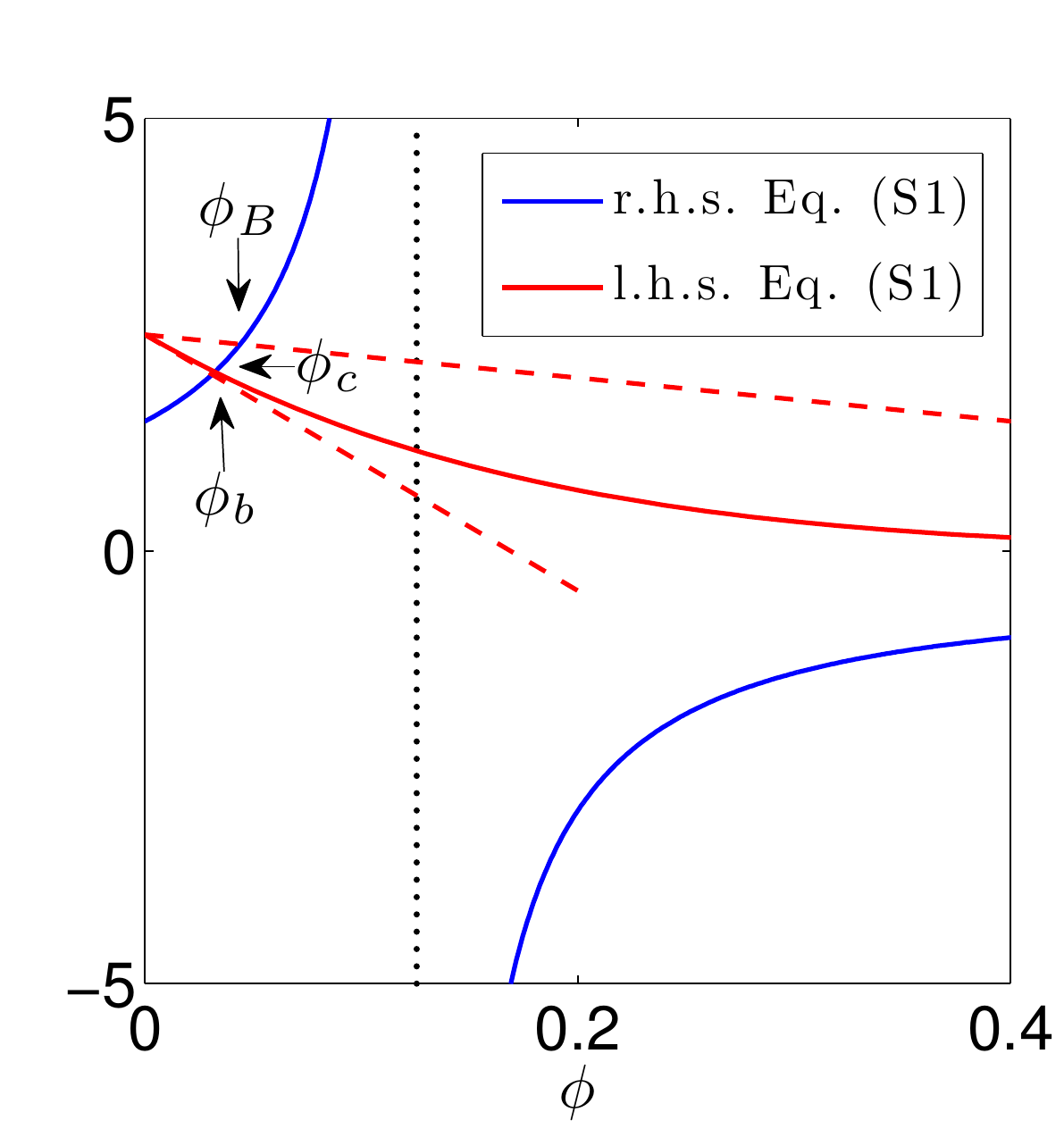}  
\caption{Bounds for the  threshold $\phi_c$.
The threshold itself is the intersection between the curves defined by the l.h.s.\  (continuous red line) and r.h.s.\ (continuous blue line) of Eq.~(\ref{eq:phiB}). This threshold is upper bounded by the intersection $\phi_B$ of the straight line $y = c ( 1- \phi)$ (top dashed 
 line) and the upper blue curve. The threshold is also lower bounded by the intersection $\phi_b$ of the straight line $y=(2c-\langle k^3 \rangle/\langle k \rangle+1)\phi +c$ (bottom dashed 
 line) and the upper blue curve.  The  dotted 
 line is the asymptote defined by  $\phi_0$.
Note that these bounds are well defined since  $(1-\phi)G_1'(1-\phi)$ is a convex function of $\phi$ in the interval (0,1), which follows from the properties of the generating function $G_1$, and this guarantees that the continuous red line will always be within the two dashed lines.
}\label{fig:theory}
\end{figure}

\bigskip
\noindent 
We consider the observability transition threshold $\phi_c$ under the assumption that the network has a giant component, which is equivalent to 
the condition $c\equiv G_1'(1)>1$. We start with Eq.\ (4) in the form
\begin{equation}
\label{eq:phiB}
(1-\phi)G_1'(1-\phi) = \frac{c-1}{1-\phi c - \phi (1-\phi) c^2}.
\end{equation}
We first note that the r.h.s.\ of this equation defines two curves separated by the asymptote at the root $\phi_0 = (1+c-\sqrt{(c+3)(c-1)})/2c$ of the denominator,  which is entirely determined by $c$ (see Fig.~\ref{fig:theory}). The l.h.s.\ depends on the generating function, but has three general properties: 
(1) it takes the value $c$ when $\phi=0$;  (2) it takes the value $0$ when $\phi=1$; and (3) it is bounded from above by $c(1-\phi)$ for any given $\phi \in (0,1)$. 

\begin{figure}
\centering
\includegraphics[scale = 0.6]{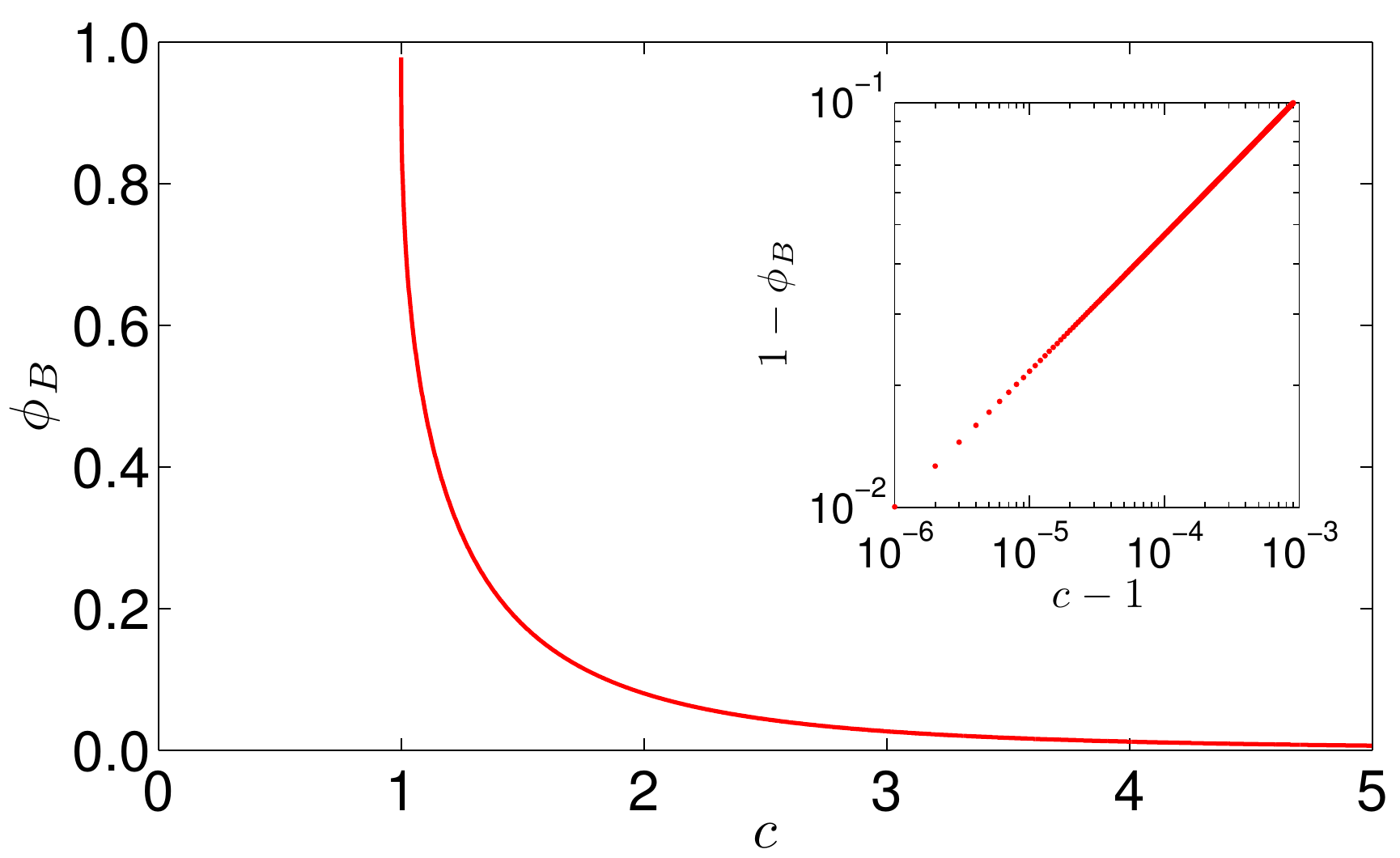}
\caption{Upper bound $\phi_B$ as a function of $c$.  The rapid decrease of $\phi_B$  as $c$ increases indicates that the threshold $\phi_c$ itself is necessarily small unless $c$ is small.
In fact, the threshold $\phi_c$  can be arbitrarily small for any given $c$ 
but is necessarily strictly positive when the third moment of the degree distribution is finite (see~Figs.~\ref{fig:theory} and \ref{fig:phiBphic}).  The inset shows that $\phi_B$ goes to $1$  as $|1-\phi_B| \sim |c-1|^{1/3}$ when $c \to 1^+$. 
}\label{fig:phiB}
\end{figure}

We can therefore calculate an upper bound for $\phi_c$ defined by the intersection between the r.h.s.\ of Eq.~(\ref{eq:phiB}) and the straight line $y = c ( 1-\phi)$, as indicated in Fig.~\ref{fig:theory}. We  denote this upper bound by $\phi_B$ and note that it depends on the degree distribution only through $c$. This upper bound converges to zero quickly  as $c$ increases (Fig.~\ref{fig:phiB}), which shows that the dependence of the threshold $\phi_c$ on higher moments of the degree distribution is necessarily small  when $c$ is not too small.  We recall that  $c= \frac{\langle k^2\rangle - \langle k\rangle}{\langle k\rangle}$ and hence the upper bound $\phi_B$  depends only on the the first two moments.  On the other hand, the upper bound  $\phi_B$ increases to $1$ as $c$ decreases to $1$. In the small-$c$ limit, we can easily derive the scaling $|1-\phi_B| \propto |1-c|^{1/3}$  by expanding Eq.~(\ref{eq:phiB}) around  $c=1$. This scaling is illustrated numerically in the inset of Fig.~\ref{fig:phiB}. 

We also calculate a lower bound for  $\phi_c$, which we denote by  $\phi_b$. This lower bound is  defined by the intersection between the upper curve defined by the r.h.s.\ of Eq.~(\ref{eq:phiB}) and the line tangent to the curve $(1-\phi)G_1'(1-\phi)$ at $\phi=0$ (see Fig.~\ref{fig:theory}). The slope of this tangent is given by
\begin{equation}
 \frac{d}{d \phi} [(1-\phi)G_1'(1-\phi)] \Big| _{\phi=0} = -G_1'(1) - G_1''(1) = 2c - \frac{\langle k^3\rangle }{\langle k\rangle } + 1.
\end{equation}
Therefore, given $c$ for a degree distribution, the ratio $\frac{\langle k^3\rangle }{\langle k\rangle }$ determines the lower bound for $\phi_c$. 
Note that an increase of this ratio causes a decrease of the lower bound $\phi_b$.

\begin{figure}
\centering
\includegraphics[scale = 0.6]{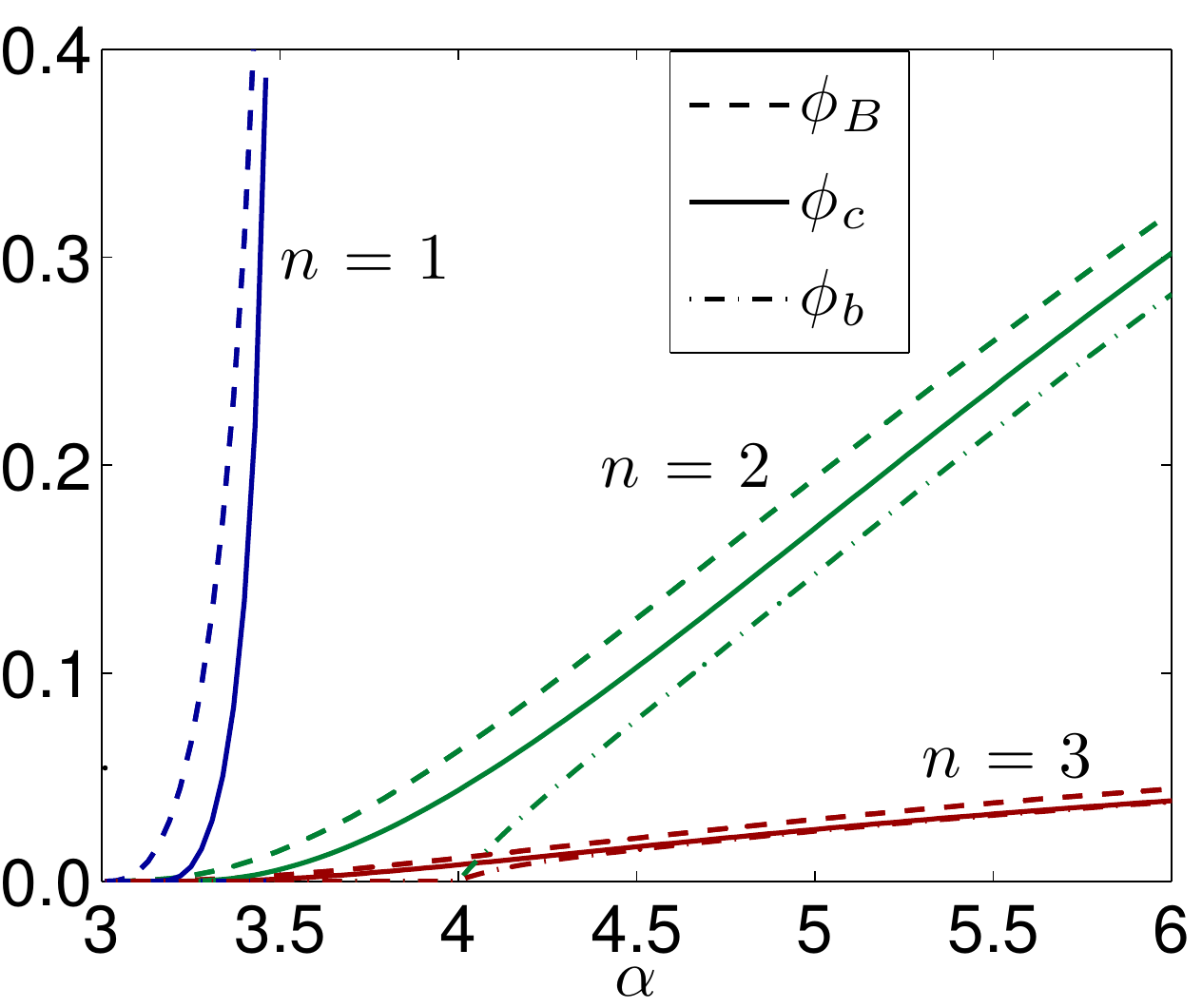}
\caption{Upper bound $\phi_B$, 
lower bound  $\phi_b$, 
and actual threshold $\phi_c$  
as functions of the exponent $\alpha$ for  power-law degree distributions. The different colors correspond to 
degree distributions with different minimum degree $n$.
In all cases the threshold $\phi_c$ increases as the scaling exponent $\alpha$ increases---i.e., as the degree distribution becomes less heterogeneous. Note that the lower bound  $\phi_b$ is zero for $\alpha\le4$ because the third moment of the degree distribution diverges there.
}\label{fig:phiBphic}
\end{figure}

For networks with a power-law degree distribution  $P(k) \propto k^{-\alpha}$ for $k\ge 1$, the parameter $c$ diverges if $\alpha \le 3$, indicating that the threshold $\phi_c$ is zero in this range. However, a nonzero threshold exists when the exponent $\alpha$ is larger than $3$, insofar as $c>1$. To further illustrate the role of higher moments, we consider power-law networks with minimum degree $n$ \cite{snewman_book}: 
\begin{equation}
P(k)=
\begin{cases}
0     &						\text{for }  k< n, \\
\frac{k^{-\alpha}}{\zeta(\alpha,n)}  &   \text{for }   k \ge n,
\end{cases}
\end{equation}
where $\zeta(\alpha,n)=\sum_{k=n}^{\infty} k^{-\alpha}$ is the incomplete zeta function.
The corresponding generating function is given by 
\begin{equation}
G_0(x) =\frac
{ Li_{\alpha}(x)-\sum_{k=1}^{n-1}k^{-\alpha}x^k}
{\zeta(\alpha,n)},
\end{equation}
where   $Li_{\alpha}(x) = \sum_{k=1}^{\infty}k^{-\alpha}x^{k}$ is the polylogarithm of $x$.  
The case $n=1$ corresponds to widely studied power-law networks, which have the inconvenience that $c>1$ only for $\alpha<3.47875...$ \cite{Aiello2000}.
The condition $c>1$ is satisfied for any $\alpha$ when $n\ge 2$. As illustrated in Fig.~\ref{fig:phiBphic}, 
for any minimum degree $n$ the bounds $\phi_B$ and $\phi_b$  and the threshold $\phi_c$ increase from zero as the scaling exponent $\alpha$ 
increases.

\newpage

\newpage$\phantom{.}$
\newpage
\centerline{\bf Robustness of Observability to Edge Rewiring}

\begin{figure*}[h] 
\includegraphics[width=9cm]{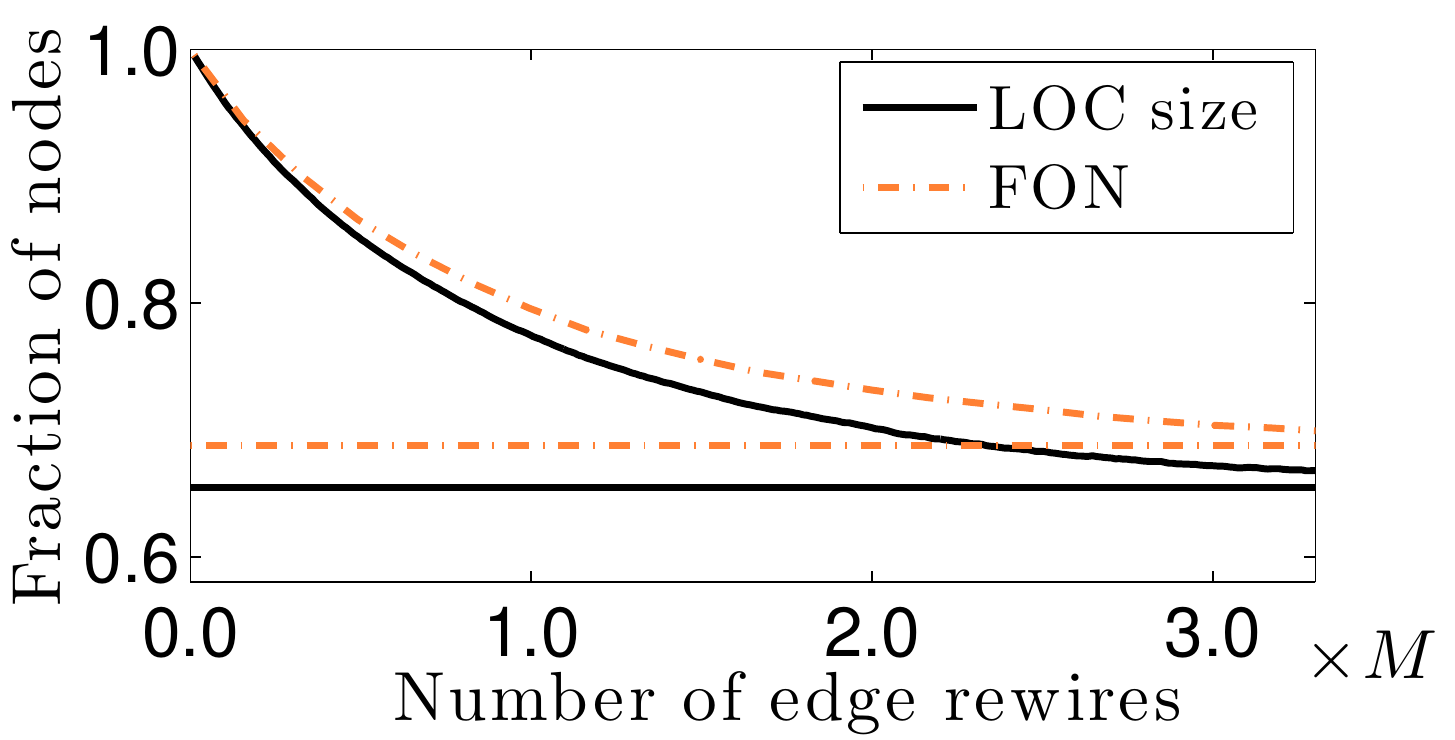}
\vspace{-0.3cm}
\caption{Robustness of LOC size and  FON to edge rewiring.
Starting with an optimal PMU placement in the Eastern North America power grid,
the curves represent the  LOC size and FON as a function of an increasing number of random edge rewires,
where $M$ is the number of edges in the network.  
The horizontal lines correspond to the random limit. 
\label{figS4}}
\end{figure*}

\bigskip
\noindent 
Both the LOC size and the FON are rather robust as functions of the number of random edge rewires constrained to keeping the network connected, as shown in  Fig.~\ref{figS4} for the Eastern North America power-grid network. In this case, the LOC size and the FON remain above $95\%$ when up to $12\%$ of the edges are rewired, 
which is nontrivial given that the network is initially equipped with an optimal 
 placement of PMUs. 
While the evolution of a power grid is 
not merely defined by edge rewiring,
this robustness corroborates the conclusion
 that, once optimal, the PMU placement will remain relatively close to optimal as 
power lines are rewired. 
The  actual planned evolution of the Eastern North America power grid is considered in  Fig.\ 3(b) of  the paper.
\\

\newpage$\phantom{.}$
\noindent
\centerline{\bf  Depth$-2$ Observability}

\begin{figure}[t!]
\includegraphics[width=10cm]{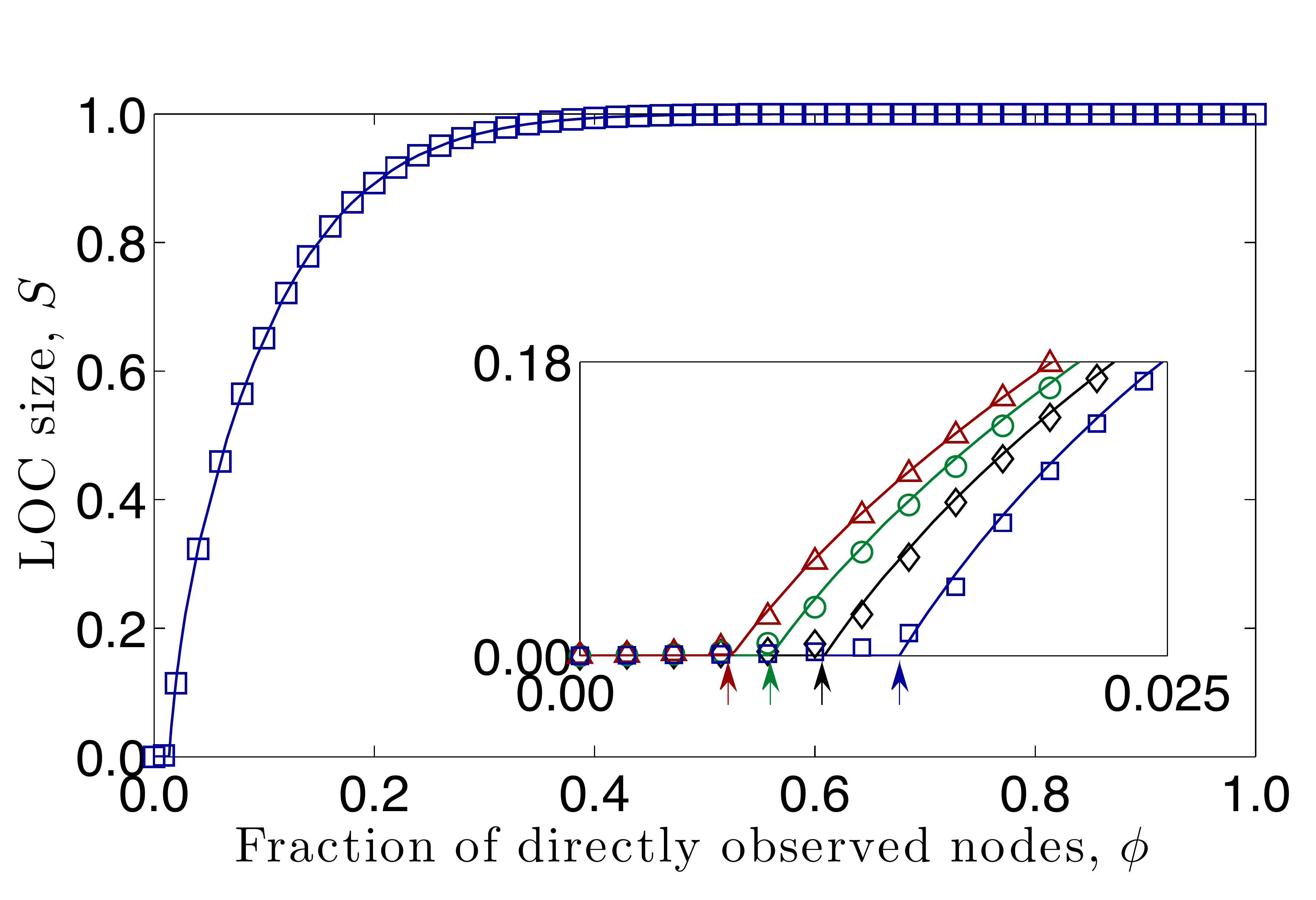}
\caption{Network observability transitions for depth-2. LOC size as a function of the fraction of directly observable nodes $\phi$ in configuration-model networks with average degree $3$.  The continuous lines correspond to our analytical prediction in Eq.~(\ref{loc2}) and the symbols to numerical simulations.  
Main panel: full dependence of the LOC size on $\phi$ for  homogeneous networks ($G_0(x)=x^3$, $\sigma^2=0$).
Inset: magnification around the transition for the homogeneous networks considered in the main panel ($\sigma^2=0$, blue) as well as for networks of increasing heterogeneity, defined by $G_0(x)=\frac{1}{6}x^2 + \frac{2}{3}x^3 + \frac{1}{6}x^4$ ($\sigma^2=1/3$, black), $G_0(x)=\frac{1}{3}x^2 + \frac{1}{3}x^3 + \frac{1}{3}x^4$ ($\sigma^2=2/3$, green), and
$G_0(x)=\frac{1}{2}x^2 +  \frac{1}{2}x^4$ ($\sigma^2=1$, red). In both panels, each symbol is an average over ten $10^5$-node  networks for $10$ independent random placements each. 
\label{figS1}}
\end{figure}

\bigskip
\noindent 
Our theory can be extended to different observability depths. We illustrate these results briefly for the depth-2 case, in which 
a sensor placed on a node makes the node's first {\it and} second neighbors observable. Using $r$ to denote the shortest path distance to the nearest observable node with a sensor, we say that a node is 0-observable (or directly observable) if $r=0$, is 1-observable if $r=1$, and is 2-observable if $r=2$. The probability that a node $i$ is not connected to the LOC though a randomly selected edge  $e_{ij}$ is then denoted $u$ if $i$ is 0-observable, $v$ if $i$ is 1-observable but $j$ is not 0-observable, and $s$ if  $i$ is 2-observable. 
Through an argument analogous to the one used in the paper, we consider the observability status of the first and second neighbors of node $i$ to obtain the self-consistent equations
\begin{subequations}
\begin{eqnarray}
u&=&\phi G_1 (u)+(1-\phi) G_1 (\Psi_1),\\
v&=&G_1[(1-\phi)s]+\Psi_2,\\
s&=&G_1[(1-\phi)G_1(1-\phi)]+\Psi_2+G_1(\Psi_3)-G_1[(1-\phi)s G_1(1-\phi)],
\end{eqnarray}
\end{subequations}
where 
\begin{subequations}
\begin{eqnarray}
\Psi_1&=&\phi G_1(u)+(1-\phi)v,\\
 \Psi_2&=&G_1(\Psi_1)-G_1[(1-\phi)v], \\
 \Psi_3&=&(1-\phi)[\Psi_2+s G_1(1-\phi)].
\end{eqnarray}
\end{subequations}
Summing over all 0-, 1-, and 2-observable nodes, it follows that the resulting  LOC size is
\begin{eqnarray}
S = 1-\phi G_0(u)- (1-\phi)\{G_0(\Psi_1)+G_0[(1-\phi)G_1(1-\phi)]-G_0[(1-\phi)v]&& \nonumber\\
 +G_0(\Psi_3) -G_0[(1-\phi)s G_1(1-\phi)]&&\}.
\label{loc2}
\end{eqnarray}
This is the depth-2 generalization of the result in Eq.~(5) of the paper, and is illustrated in Fig.~\ref{figS1}.

\newpage$\phantom{.}$
\newpage
\noindent
\centerline{\bf Observability in a Metabolic Network}

\bigskip
\noindent 
In a metabolic network consisting of $N$ reactions and $m$ metabolic compounds, the problem of observability can be formulated in terms of the set of $K$ reactions that need to be measured to determine the state of the entire network. The state of a metabolic network is defined by the fluxes of all of its reactions, which we represent as a vector $\mathbf{v}=(v_{j})$. In steady state, which is the most widely studied condition in experiments, the vector of fluxes satisfies
\begin{equation}
\mathbf{S} \cdot \mathbf{v} = 0,
\label{eq:sv}
\end{equation}
where $\mathbf{S}=(S_{ij})$ is the stoichiometric matrix accounting for the structure of the network. In this equation, $S_{ij}$ represents the stoichiometric coefficient of the $i$th metabolic compound in the $j$th reaction, and $v_j$ is the flux of the $j$th reaction.  Because the number of reactions is generally larger than the number of metabolites, Eq.~(\ref{eq:sv}) is underdetermined.  
The problem of optimal measurement placement for complete observability is then reduced to the identification of the smallest set of  reaction fluxes that need to be measured in order to determine  $\mathbf{v}$ uniquely given the constraints imposed by (\ref{eq:sv}). This number is simply $N-r$, where $r$ is the rank of the $m\times N$ matrix $\mathbf{S}$. 

The reactions can be categorized into two classes, one consisting of a total of $N_1$ biochemical and transport reactions and the other consisting of $N_2$ exchange reactions (fictitious reactions representing the transport of metabolic species across the system boundary). The fluxes of the reactions in the first class (but not in the second) can be determined experimentally.   
The columns of matrix $\mathbf{S}$ corresponding to exchange reactions form a set of linearly independent vectors, meaning that the optimal number of measurements to determine $\mathbf{v}$ continues to be $N-r$ even if the measurements are limited to the set of $N_1$ biochemical and transport reactions.

A related question concerns the observability of the network upon random placement of measurements. This problem can be formulated as follows.  Assume we measure $p$ reaction fluxes and rewrite Eq.~(\ref{eq:sv}) as
\begin{equation}
\left[
\begin{array}{cc}
\mathbf{S} _1 & \mathbf{S} _2
\end{array} \right]
 \cdot 
 \left[ 
\begin{array}{c}
\mathbf{v}_1 \\
\mathbf{v}_2\\
\end{array} \right] = 0,
\label{eq:sv2}
\end{equation}
where the reactions have been re-indexed such that  $\mathbf{S}_1$ is a $m \times p$  matrix and $\mathbf{v_1}$  represents the $p$ measured reaction fluxes.  This equation can be reorganized as
\begin{equation}
\mathbf{S} _2 \cdot \mathbf{v}_2 = -\mathbf{S} _1 \cdot \mathbf{v}_1,
\label{eq:sv3}
\end{equation}
which shows that $\mathbf{v}_2$ is uniquely determined iff the rank of matrix $\mathbf{S}_2$, denoted $r_2$, equals the number 
of variables $N-p$  in $\mathbf{v}_2$. When $r_2<N-p $,
we may still partially determine $\mathbf{v}_2$ using 
Gauss-Jordan elimination
to obtain a reduced row echelon form of $\mathbf{S}_2$ \cite{matrix_book}.
We denote this row echelon form  as $\mathbf{S}_2'$.  If any row of $\mathbf{S}_2'$ contains only one nonzero element, then the reaction corresponding  to this nonzero element is uniquely determined. The uniquely determined elements  of  $\mathbf{v}_2$ and corresponding columns of  $\mathbf{S}_2$ can be moved to the right side of Eq.~(\ref{eq:sv3}), which we implement by redefining the corresponding matrices and vectors. 
Numerically, 
we randomly select 
reactions that are not yet uniquely determined
among the $N_1$ biochemical and transport reactions  in the network and we repeat this process iteratively by updating $\mathbf{v}_{1}$,  $\mathbf{v}_{2}$, $\mathbf{S} _{1}$, and $\mathbf{S} _{2}$ at each step.
The whole processes is computationally efficient because, once the Gauss-Jordan elimination has been implemented for the initial matrix $\mathbf{S}_2$,  
the updates of matrix $\mathbf{S}_2$ are kept in the reduced row echelon form as we remove the columns associated with uniquely determined reactions.

As an application, we consider the most complete reconstruction of the human metabolic network \cite{metabolic_network}, 
which has $N_1=3,338$ biochemical and transport reactions, $N_2=404$ exchange reactions, $m=2,766$ metabolic compounds, and is the largest metabolic network available in the literature. The rank of the resulting $2,766\times 3,742$ matrix $\mathbf{S}$ is $r=2,674$, meaning that the entire network is observable by measuring $1,068$ properly selected reactions; this corresponds to $28.5\%$ of all  
reactions, hence a fraction comparable to the one found for the optimal placement of PMUs in power grids. 
To determine how observability changes as a function of the number of randomly selected  reactions  measured, we calculate the FON and LOC size for the reactions regarded as nodes and the metabolic compounds as edges.

\begin{figure} [h!]
\centering
\includegraphics[scale = 0.65]{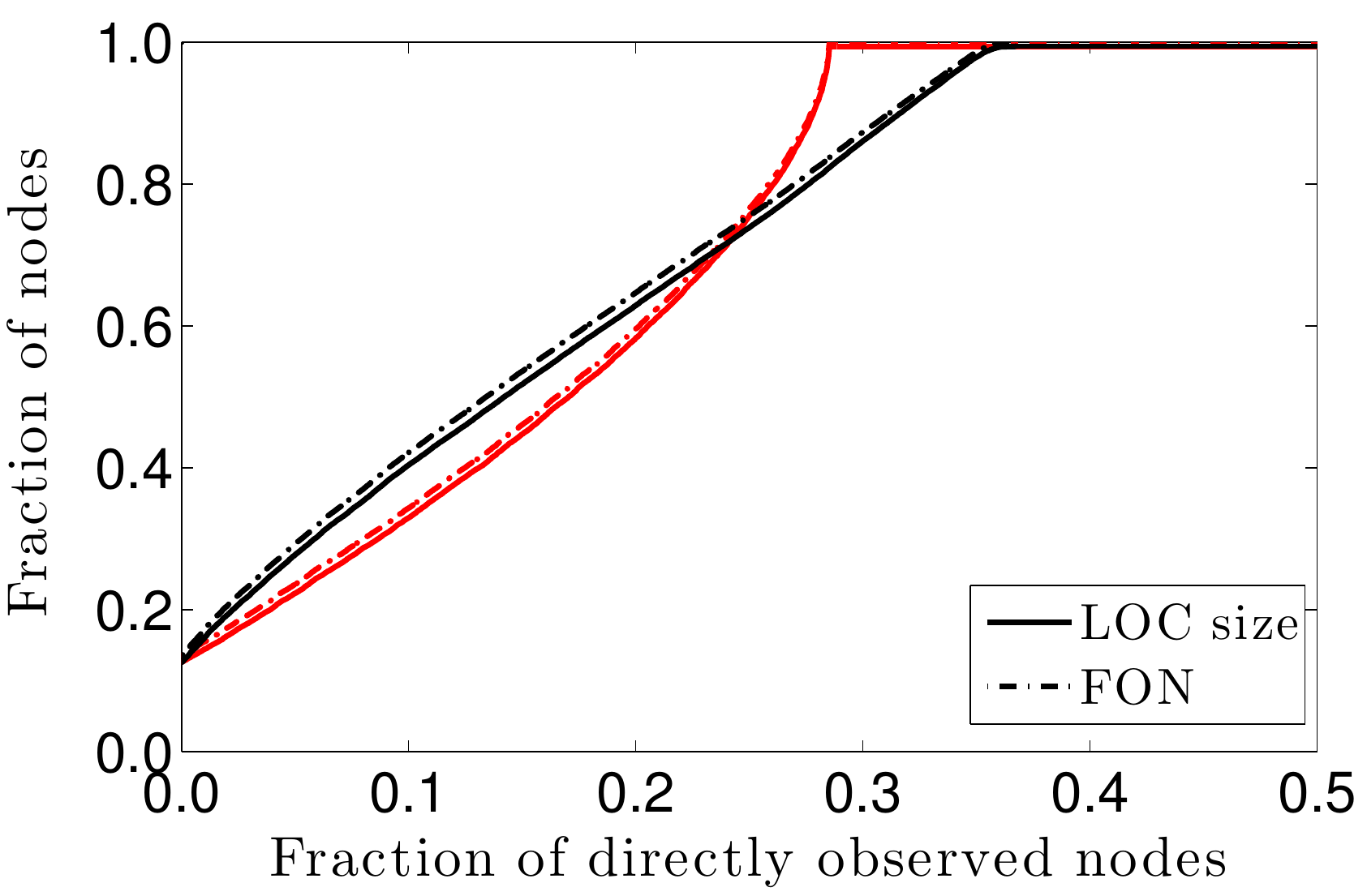}
\vspace{-0.3cm}

\caption{Observability of the human metabolic network. 
The  LOC size and FON  as a function of the fraction 
of reactions measured for measurements placed on the full network (black) and on 
the optimal set (red).
The lines represent an average over $100$ independent realizations in which
the  measured reactions are selected randomly, as described in the text.
}
\label{fig:Q1fig1}
\end{figure}

As shown in Fig.~\ref{fig:Q1fig1}, for the human metabolic network, there is no significant difference between the LOC size and the FON, and both grow rapidly with the fraction of measured reactions. This holds true both when the random placement of measurements is performed on the full network of internal and transport reactions (Fig.~\ref{fig:Q1fig1}, black lines) and when the random placement is limited to the optimal set (Fig.~\ref{fig:Q1fig1}, red lines).  These properties are due to the presence of metabolic compounds in the network, such as ATP, which are involved in a large number of reactions.  
Note that even when no reactions are measured the LOC size and FON are nonzero and this is so because a total of $13.5\%$ of the reactions are uniquely determined by Eq.~(\ref{eq:sv}) alone (these correspond to reactions that are inactive in any steady state \cite{joo_sang2012}). Different from the power grid case, the transition starts at zero even if we ignore these reactions upfront.

The difference between random placement on  the optimal set or on the full network is relatively small in the metabolic case, with the full observability achieved when approximately  $28.5\%$ and $36.2\%$ of the reactions are measured, respectively. Surprisingly, for a small number of measurements, the LOC size and FON are larger for random placements on the entire network than on the optimal set; in the latter case,  the LOC size and FON increase sharply as the number of measurements goes from nearly all to all reactions in the optimal set. This is the case because for a reaction to be indirectly observable it usually requires multiple of its network neighbors to be directly observable and, for measurement placements limited to the optimal set, this condition is only rarely satisfied until most reactions in the set have been measured.

\begin{thebibliography}{}
\bibitem{comment_1}
We do not discriminate zero-injection nodes as this makes the results more broadly applicable.
\end{thebibliography}
\end{document}